\documentclass[pra,aps,twocolumn,longbibliography,nobalancelastpage, superscriptaddress]{revtex4-1}
\usepackage[utf8]{inputenc}
\usepackage[T1]{fontenc}
\usepackage{amsmath}

\usepackage[dvipsnames]{xcolor}
\usepackage{bbold}
\usepackage{chemformula}
\colorlet{myPurple}{blue!40!red}
\colorlet{myPurplee}{blue!10!red}
\colorlet{myCyan}{cyan!60!gray}
\colorlet{myRed}{red!66!black}
\usepackage{tikz}
\usepackage{pgfplots}
\usepackage{float}
\usepackage{geometry}
\pgfplotsset{compat=1.14}
\usepackage[colorlinks=true,citecolor=myRed,urlcolor=myRed,linkcolor=myRed]{hyperref}
\usepackage[normalem]{ulem}
\usepackage{exscale}
\usepackage{bbm}
\usepackage{graphicx}
\usepackage{amsmath}
\usepackage{latexsym}
\usepackage{amsfonts}
\usepackage[separate-uncertainty=true,multi-part-units=single]{siunitx}
\usepackage{amssymb}
\usepackage{times}
\usepackage[T1]{fontenc}
\usepackage{amsthm}
\usepackage{enumerate}
\usepackage{bbold}
\usepackage{color}
\usepackage{nicefrac}
\usepackage{soul}
\definecolor{olivegreen}{HTML}{3C8031}
\newcommand{\sket}[1]{{\ensuremath{\lvert#1\rangle}}}
\newcommand{\lket}[1]{{\ensuremath{\left\lvert#1\right\rangle}}}
\newcommand{\ket}[1]{\if@display\lket{#1}\else\sket{#1}\fi}

\newcommand{\sbra}[1]{{\ensuremath{\langle#1\rvert}}}
\newcommand{\lbra}[1]{{\ensuremath{\left\langle#1\right\rvert}}}
\newcommand{\bra}[1]{\if@display\lbra{#1}\else\sbra{#1}\fi}

\newcommand{\sbraket}[2]{{\ensuremath{\langle#1\rvert#2\rangle}}}
\newcommand{\lbraket}[2]{{\ensuremath{\left\langle#1\!\left\rvert\vphantom{#1}#2\right.\!\right\rangle}}}
\newcommand{\braket}[2]{\if@display\lbraket{#1}{#2}\else\sbraket{#1}{#2}\fi}

\newcommand{\sketbra}[2]{{\ensuremath{\lvert #1\rangle\!\langle #2\rvert}}}
\newcommand{\lketbra}[2]{{\ensuremath{\left\lvert #1\right\rangle\!\!\left\langle #2\right\rvert}}}
\newcommand{\ketbra}[2]{\if@display\lketbra{#1}{#2}\else\sketbra{#1}{#2}\fi}
\usepackage{tcolorbox}
\usepackage{mathtools}

\usepackage{tikz}
\usepackage{lipsum}
\theoremstyle{plain}

\usepackage[font=small,labelfont=bf,justification=justified,format=plain]{subcaption}

\usepackage{graphicx}
\usepackage{bm}
\usepackage{dsfont}
\usepackage{tikz}
\usepackage[T1]{fontenc}
\usepackage{amsthm}
\usepackage{array}
\usepackage{amssymb}
\usepackage{amsfonts}
\usepackage{cancel}
\usepackage[toc,page]{appendix}
\usepackage{multirow}
\usepackage{color}
\usepackage{calrsfs}
\usetikzlibrary{backgrounds,decorations.pathreplacing,calc}

\newcommand{\sinc}{\textrm{sinc}}
\usepackage{tkz-euclide}
\usepackage{tcolorbox}
\usepackage{ragged2e}
\DeclareCaptionJustification{justified}{\justifying}
\DeclareMathAlphabet{\mathcal}{OMS}{cmsy}{m}{n}
\usepackage{thmtools}
\usepackage{thm-restate}

\begin{document}

\title{Open-Path Detection of Organic Vapors via Quantum Infrared Spectroscopy}
\author{Simon Neves}\thanks{Corresponding author: simon.neves@laposte.net}
\affiliation{Group of Applied Physics, University of Geneva, rue de l'Ecole-de-Médecine 20, 1205 Geneva, Switzerland}

\author{Adimulya Kartiyasa}
\affiliation{Group of Applied Physics, University of Geneva, rue de l'Ecole-de-Médecine 20, 1205 Geneva, Switzerland}

\author{Shayantani Ghosh}
\affiliation{Group of Applied Physics, University of Geneva, rue de l'Ecole-de-Médecine 20, 1205 Geneva, Switzerland}

\author{Geoffrey Gaulier}
\affiliation{Group of Applied Physics, University of Geneva, rue de l'Ecole-de-Médecine 20, 1205 Geneva, Switzerland}

\author{Luca La Volpe}
\affiliation{Group of Applied Physics, University of Geneva, rue de l'Ecole-de-Médecine 20, 1205 Geneva, Switzerland}

\author{Jean-Pierre Wolf}
\affiliation{Group of Applied Physics, University of Geneva, rue de l'Ecole-de-Médecine 20, 1205 Geneva, Switzerland}

\date{\today}

\begin{abstract}
In recent years, quantum Fourier transform infrared (QFTIR) spectroscopy emerged as an alternative to conventional spectroscopy in the mid-infrared region of the spectrum. By harnessing induced coherence and spectral entanglement, QFTIR offers promising potential for the practical detection of organic gasses. However, little research was conducted to bring QFTIR spectrometers closer to domestic or in-field usage.  In this work, we present the first use of a QFTIR spectrometer for open-path detection of multiple interfering organic gases in ambient air. The accurate identification of mixtures of acetone, methanol, and ethanol vapors is demonstrated with a QFTIR spectrometer. We achieved this breakthrough by building a nonlinear Michelson interferometer with 1.7m-long arms to increase the absorption length, coupled with analysis techniques from differential absorption spectroscopy. The evolution of different gasses' concentrations in ambient air was measured through time. These results constitute the first use-case of a QFTIR spectrometer as a detector of organic gasses, and thus represent an important milestone towards the development of such detectors in practical situations. 
\end{abstract}

\maketitle

\large
\noindent\textbf{Introduction}
\normalsize
\medskip

\noindent Infrared absorption spectroscopy is a particularly flexible tool for the detection of volatile organic compounds (VOCs), with applications ranging from pollution monitoring~\cite{IndustrialPollution}, to breath-analysis-based diagnoses~\cite{BiomarkersReview}. In particular, quantum-cascade laser enables fast and remote VOC detection, by leveraging dual-comb spectroscopy \cite{DualCombSpectroscopy2016}. Yet the accessible bandwidth is limited to a few hundred of nanometers, and high-sensitivity MIR detectors often require cryogenic cooling \cite{MIRdetectonReview}, which limits the use of this method for tracking traces of multiple VOCs in the field. More conventional thermal sources provides a broader spectral band via Fourier transform infrared (FTIR) spectroscopy \cite{FTIRBook}, but the low brightness and directivity of such sources~\cite{MIRclassicalReview} limit the sensitivity, making the use of classical FTIR spectroscopy impractical for remote and real-time VOC tracking.

Recently, significant advances in infrared spectroscopy were achieved, by using entangled pairs of photons generated in nonlinear crystals via spontaneous parametric down-conversion (SPDC) \cite{kalashnikov2016infrared,paterova2018measurement,lindner2020fourier,mukai2021quantum,lindner2021nonlinear,arahata2022wavelength,mukai2022quantum,lindner2023high,ultrabroadbandQFTIR,cardoso2024methane}. Typically, samples were probed with photons in the MIR region, while their twins were measured in the near-infrared (NIR) or visible region. By leveraging induced coherence in nonlinear interferometers \cite{zou1991induced,lemos2014quantum} and quantum correlations in the photons' spectra, the MIR spectral information was retrieved by measuring the NIR/visible photon only, therefore bypassing shortcomings of MIR detectors and sources. This principle was applied in a nonlinear Michelson interferometer to demonstrate the first Quantum FTIR (QFTIR) spectrometer \cite{lindner2020fourier}. Additionally, by optimizing group-velocity matching (GVM) during the SPDC process \cite{lindner2021nonlinear,GVM2019}, and using a chirped nonlinear crystal \cite{ultrabroadbandQFTIR}, photons were entangled on broad spectral bands, dwarfing the spectral bandwidths accessible by commercial off-the-shelf MIR sources \cite{MIRclassicalReview}. Finally, by placing the crystal in an optical resonator, it was shown that QFTIR spectroscopy could be achieved using a low-power diode laser pump \cite{lindner2023high}. These advances make photonic quantum sensing a promising candidate for cheap and practical detectors of remote VOC traces, based on MIR absorption spectroscopy.

Despite these encouraging results, little work was carried out to showcase QFTIR spectroscopy's potential as a detector of VOC vapors in practical outdoor settings. As of today, only methane \ch{CH4} and nitrous oxide \ch{N2O} were measured in gaseous form~\cite{kalashnikov2016infrared,lindner2021nonlinear,lindner2023high,cardoso2024methane}, solely as reference gasses in sealed gas cells, over short distances, and without overlapping absorption bands. To reveal its full capabilities, QFTIR spectroscopy must be tested under conditions closer to real-world scenarios~\cite{bongs2023quantum}, including long-distance measurements, ambient air conditions, multiple interfering gasses, and uncontrollable fluctuation factors.

Our work intends to bridge this gap, by demonstrating the first open-path detection of multiple VOC vapors in ambient air via QFTIR spectroscopy. We built a nonlinear Michelson interferometer with $\SI{1.7}{m}$-long arms, thus maximizing the absorption of gasses spread over the open path, and enabling unprecedented sensitivity for a QFTIR spectrometer. The nonlinear crystal was designed to fulfill the GVM condition, thus generating entangled photons over a large spectral band, from $\SI{2700}{\per\centi\meter}$ to $\SI{3100}{\per\centi\meter}$ in the MIR. We probed mixtures of VOC vapors from 3 different species, all of which display a strong absorption over this spectral range, namely methanol, ethanol, and acetone. To address slow spectral fluctuations and interfering absorption spectra, we leverage \textit{differential absorption spectroscopy} \cite{DOASBook}, a technique that is typically used in the ultraviolet region. This approach allowed us to accurately identify different VOC mixtures by recognizing their absorption cross-sections. Finally, we tracked the VOCs' concentration in ambient air throughout the evaporation of methanol and acetone. As the first demonstration of a QFTIR spectrometer for VOC detection in ambient air, these results mark a significant step towards the practical application of this novel quantum sensor.\newpage

\large
\noindent\textbf{Quantum Spectroscopy for Gas Detection}
\medskip
\normalsize

\noindent Absorption spectroscopy is a powerful method for remote and non-invasive gas detection \cite{DOASBook}. When interacting with electromagnetic radiation, gasses absorb a fraction of the field, with an amplitude that depends on the field's frequency, as well as the concentration and nature of the different interacting compounds. The absorption spectrum of a mixture of $N$ gasses is governed by Beer-Lambert's (BL) law: 
\begin{equation}\label{eq:BeerLambert}
       \mathcal{A}(\nu) = -\ln\bigl(T(\nu)\bigr) = L\cdot\sum_{k=1}^Nc_k\cdot\sigma_k(\nu),
\end{equation}
where $\mathcal{A}(\nu)$ is the absorbance, $T(\nu)$ is the frequency-dependent transmission, $L$ the interaction length, $\sigma_k(\nu)$ the cross-section of the $k$-th gas in the mixture, and $c_k$ is its average concentration over the interaction length. Much like a fingerprint, the cross-section of each gas is unique. In principle, BL law can therefore be used to identify gasses present in a mixture, and evaluate their average concentrations $c_k$. The successful identification of the gasses relies heavily on the sensitivity of the detector used to measure the transmission $T(\nu)$, as well as its spectral bandwidth and resolution.\\

\begin{figure*}
    \centering
\includegraphics[width=178mm]{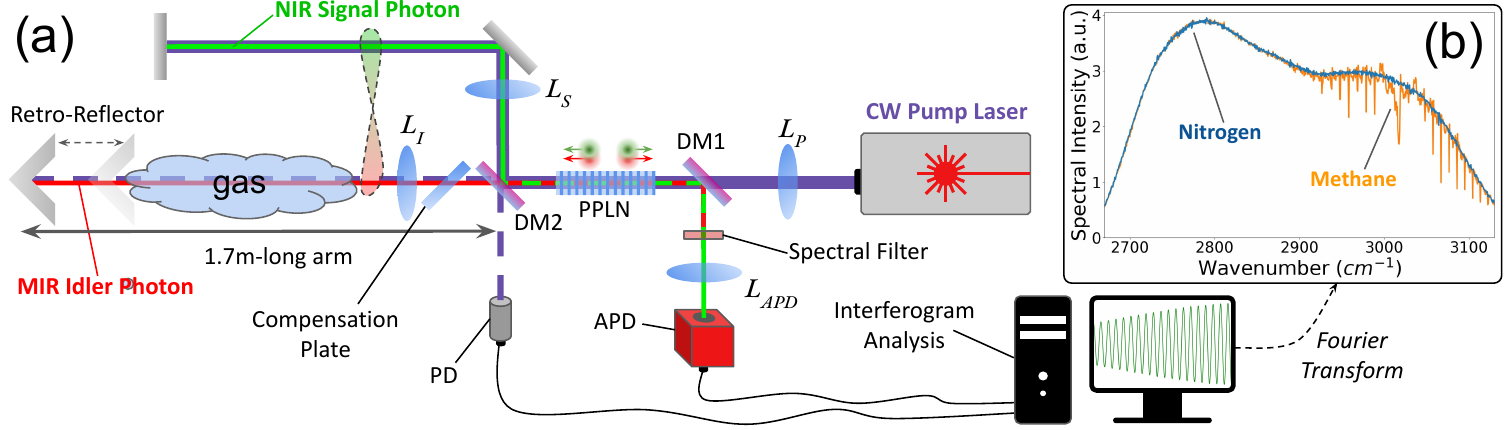}
    \caption{(a) Nonlinear Michelson interferometer used for open-path QFTIR spectroscopy. (b) Example spectra recorded when placing a gas cell filled with nitrogen (blue) or methane (yellow) in the MIR arm.}
    \label{fig:XPSetup}
\end{figure*}

Quantum FTIR spectrometers have recently exhibited remarkable performances on these benchmarks, by leveraging spectral entanglement and nonlinear interference \cite{lindner2020fourier,mukai2021quantum,lindner2021nonlinear,arahata2022wavelength,mukai2022quantum,lindner2023high,ultrabroadbandQFTIR,cardoso2024methane}. Via this method, broad spectral bands in the MIR region can be reconstructed through the detection of visible or NIR radiation only. A QFTIR spectrometer typically consists of a nonlinear Michelson interferometer, such as the one shown in~Fig.~\ref{fig:XPSetup}. A periodically-poled nonlinear crystal (PPLN in the figure) is pumped by a narrow-line pump laser, generating spectrally-entangled signal and idler photons via SPDC. This conversion process occurs with low probability, either on the pump's first passage or its second passage, after all modes have traveled through the interferometer and been reflected back into the crystal. The overlap of all beams in the interferometer erases the \textit{which-way} information, making it impossible to know whether the photons were emitted on the first or second passage. By \textit{induced coherence}~\cite{zou1991induced}, an interference occurs on the signal when the idler's path-length changes. The Fourier transform of the resulting interferogram gives the spectral intensity $I(\nu_i)$ of the idler photon, including the absorption of a potential sample placed on its path~\cite{paterova2018measurement}: 
\begin{equation}\label{eq:IdlerIntensity}
    I(\nu_i) \propto F(\nu_p-\nu_i,\nu_i)\cdot T(\nu_i) = S(\nu_i) \cdot T(\nu_i),
\end{equation}
where $\nu_i$ is the idler's frequency, $\nu_p$ is the frequency of the pump, $F(\nu_s,\nu_i)$ is the \textit{joint spectral intensity} (JSI) giving the spectral state of photons, and $T$ is the transmission of the sample. Note here that $\nu_s= \nu_p-\nu_i$ is the signal's frequency, assuming the pump laser is monochromatic. In principle, the evaluation of the absorbance spectrum $\mathcal{A}(\nu_i)$ of a gas sample comes naturally from the following relation:
\begin{equation}\label{eq:TransmExact}
    \mathcal{A}(\nu_i) = -\ln\bigl(T(\nu_i)\bigr) = \ln\bigl(I_0(\nu_i)/I_1(\nu_i)\bigr),
\end{equation}
where $I_1(\nu)$ is the spectral intensity measured when placing the gas sample in the idler's path, and $I_0(\nu)$ is the reference spectral intensity, measured when removing the sample.\\

In a typical QFTIR spectrometer, the nonlinear crystal is carefully designed for collinear emission of a signal photon in the visible or NIR region of the spectrum, and an idler photon in the MIR. In this way, gasses are probed in the MIR region, whereas only the visible-NIR photon is detected, thus bypassing the limitations of MIR detectors \cite{lemos2014quantum,paterova2018measurement}. In addition, photons can be entangled on large spectral bands, by optimizing the GVM during the SPDC process~\cite{GVM2019,lindner2021nonlinear}, or by designing chirped nonlinear crystals \cite{ultrabroadbandQFTIR}. In principle, this large band allows to collect enough information on the absorption spectrum of a gas mixture in the MIR, in order to identify compounds such as VOCs via eq.~\ref{eq:BeerLambert}. Still, significant challenges need to be addressed for performing such measurements in practice, in ambient air and over time. Firstly, the ability of QFTIR to measure trace gasses remotely and in low concentrations remains to be demonstrated, which typically requires an increase in sensitivity through elongation of the interaction length with the gasses. Secondly, slow fluctuations commonly occur in the JSI, typically caused by temperature or alignment variations in the nonlinear crystal. Finally, it is generally impractical to isolate each absorber in a different spectral band in order to identify them separately, such as \ch{CH_4} and \ch{N_2O} in \cite{lindner2023high}. In the field, one has to deal with multiple interfering absorbers in the same spectral bands (including undesirable compounds), and with slow transmission fluctuations induced by diffusion and temperature variations in ambient air. Our work addresses these limitations, in order to provide the first practical use-case of QFTIR spectroscopy for the remote detection of multiple interfering VOCs, in ambient air, and over time.\\

\large
\noindent\textbf{Open-Path QFTIR Spectrometer}
\normalsize
\medskip

\noindent Our QFTIR spectrometer is made of a nonlinear Michelson interferometer as displayed in Fig.~\ref{fig:XPSetup}. The pump is a tunable continuous-wave (CW) Ti:sapphire laser (Matisse-CR from \textit{Spectra-Physics}) with stabilized wavelength, and a narrow linewidth of $<\SI{1}{\mega\hertz}$. For most of our experiments, we set the wavelength to $\SI{800}{\nano\meter}$ and the output power to $\SI{1}{W}$. This laser pumps a periodically-poled \ch{MgO}-doped Lithium Niobate nonlinear crystal (PPLN). This PPLN has a poling period of $\SI{22.1}{\micro\meter}$, a length of $\SI{10}{\milli\meter}$, and is stabilized in temperature by an oven. At a temperature of $\SI{74}{\degree\C}$, energy conservation and type-0 quasi-phase matching through the SPDC process impose the emission of a signal-photon around $\SI{1040}{\nano\meter}$ in the NIR, and an idler-photon around $\SI{3500}{\nano\meter} \simeq \SI{2900}{\per\centi\meter}$ in the MIR. We measure a signal power of $\SI{12.3}{\nano\watt}\pm\SI{0.2}{\nano\watt}$. Lenses $L_P$, $L_S$, $L_I$ have a $\SI{100}{\milli\meter}$ focal length, ensuring a weak focus of the different modes in the crystal. In that case~\cite{lindner2021nonlinear, GVM2019}, the GVM condition is close-to-fulfilled, so the photons' bandwidth is close-to-maximized. Thanks to the pump's narrow linewidth, the photons are maximally-entangled in spectrum. All beams travel a $L=\SI{1.7}{\meter}$-distance, and gas samples are probed on the idler's open-path. All beams are then reflected back in the crystal, producing the nonlinear interference. A retroreflecting hollow roof prism mirror is used for this purpose in the MIR arm, in order to facilitate the alignment and stabilize the setup over long time-spans. After the dichroic mirror DM1 and spectral filter, only the NIR signal is focused by the lense $L_{APD}$ on the avalanche photodiode (APD, A-CUBE-S500-10 from \textit{laser Components}). As described in the previous paragraph, the APD detects an interference of the signal when the idler's path-length moves, over a $\Delta x = \SI{9}{\milli\meter}$ available distance, at a $\SI{4}{\milli\meter\per\second}$ speed. The APD's signal passes through a $\SI{0.3}{\kilo\hertz}-\SI{10}{\kilo\hertz}$ electronic band-pass filter for noise cancellation, and is then sampled at a $\SI{100}{\kilo\hertz}$ rate. The idler's path length is tracked precisely by measuring the interference of the pump beam's leak through the dichroic mirror DM2 with a reference photodiode (PD). A compensation plate is used to correct the misalignment between the MIR photon and the pump, due to DM2's dispersion. Finally, the Fourier transform of the signal's interferogram gives the idler's spectrum, including the absorption of the gas sample, as described in eq.~\ref{eq:IdlerIntensity}. In this way we measure a large $\SI{400}{\per\centi\meter}$ spectral band in the MIR (see Fig.~\ref{fig:XPSetup}.b), which contains strong vibrational-rotational absorption lines of all gasses of interest in this study. \\

We first tested our spectrometer on reference gasses made of methane \ch{CH_4} and nitrogen \ch{N_2}, in different known proportions. Typical spectra are displayed in Figure~\ref{fig:XPSetup}.b, showing methane's thin absorption lines, and the transparency of nitrogen in our detector's bandwidth. We used pure nitrogen to record the reference spectrum $I_0$ in eq.~\ref{eq:TransmExact}. As any detector, our spectrometer induces a response $H$ when recording an interferogram, which in our case takes the following form: 
\begin{equation}\label{eq:detResp}
   H(\nu) = \dfrac{2}{\Delta\nu}\sinc\biggl(\dfrac{2\pi\nu}{\Delta\nu}\biggr), 
\end{equation}
where $\Delta\nu$ is the spectral resolution. In practice, we therefore evaluate the following quantity in place of the absorbance $\mathcal{A}$: 
\begin{equation}\label{eq:TransmEval}
   \Tilde{\mathcal{A}}(\nu_i)= -\ln\Bigl(\dfrac{[I_1 * H](\nu_i)}{[I_0 * H](\nu_i)}\Bigr) \simeq -\ln\bigl([T * H](\nu_i)\bigr),
\end{equation}
where the approximation in eq.~\ref{eq:TransmEval} is implicitly made in most studies, and is discussed in the following paragraphs. We make the same assumption in this work, and verify through calibration of our detector with the reference gasses that $\Delta\nu = (\Delta x)^{-1} \simeq \SI{1.11}{\per\cm}$. This calibration was performed by fitting the experimental transmission spectrum with eq.~\ref{eq:TransmEval}, where the transmission $T$ of methane was reconstructed from BL law (eq.~\ref{eq:BeerLambert}), and the cross-section was taken from the HITRAN database \cite{Hitran2020} (See Supp. Mat. \ref{sect:methane} for details on the calibration). To derive the sensitivity of our spectrometer to methane, we probed a \ch{CH_4}-\ch{N_2} mixture containing $\SI{100}{ppm}$ (parts per million) of methane, in a $\SI{1.35}{m}$-long sealed gas-cell, at a $\SI{1}{\bar}$ pressure and $\SI{20}{\degree\C}$ temperature. By averaging the spectra over 100 scans $\Delta x$ on the idler's arm, we measured a signal-to-noise ratio (SNR) of $\approx 24$ for the characteristic $\SI{3018}{\per\cm}$-peak. This way, for methane spread over the whole $\SI{1.7}{\meter}$-arm, we estimate our detector to be sensitive to concentration changes as low as $\lesssim\SI{4}{ppm}$. This represents a sensitivity improvement of nearly two orders of magnitude compared to the previous study~\cite{lindner2021nonlinear} conducted under similar conditions, mostly attributable to the increased interaction length $L$ (see eq.~\ref{eq:BeerLambert}).\\ 

\begin{figure*}
    \centering
\includegraphics[width=178mm]{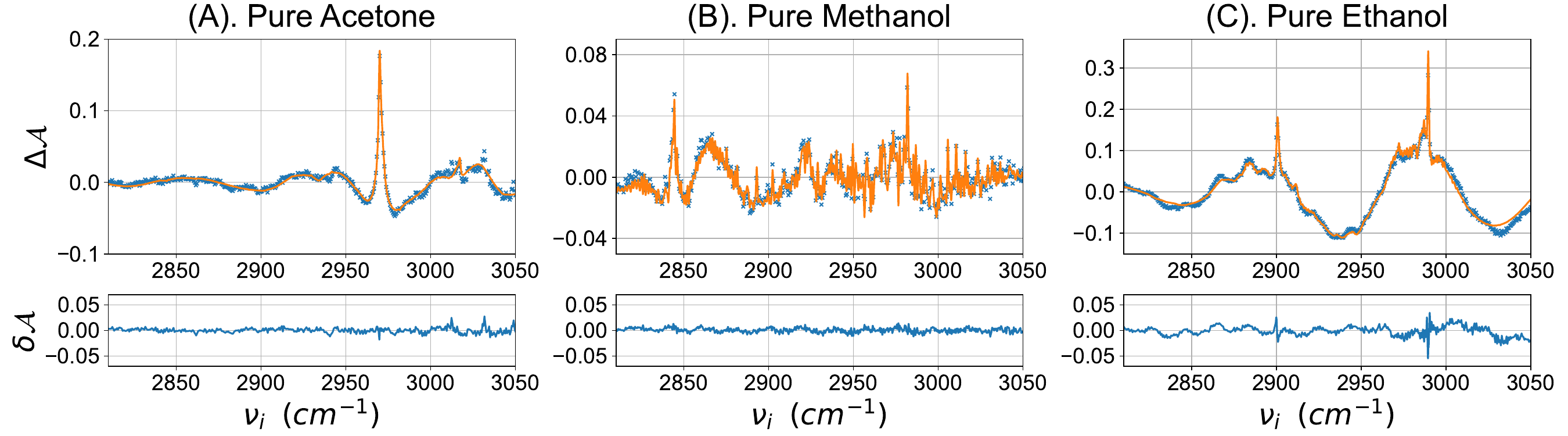}
    \caption{Differential absorbance $\Delta\mathcal{A}$ recorded by our QFTIR spectrometer, probing vapors emitted from (A) pure liquid acetone, (B) pure liquid methanol, (C) pure liquid ethanol. The experimental data (in blue) is fitted by the theoretical absorbance (in~yellow, see expression \ref{eq:diffAbsorbance}). We also display the deviation $\delta\mathcal{A}$ of experimental data from the theoretical absorbance, or residual.}
    \label{fig:CSVOC}
\end{figure*} 

\large
\noindent\textbf{VOC Detection via QFTIR Spectroscopy}
\normalsize
\medskip

\noindent We now present a first use-case of our QFTIR spectrometer for the detection of VOC vapors in ambient air. The idea is to measure the transmission $T$ (eq.~\ref{eq:TransmExact} and \ref{eq:TransmEval}) of a gas sample over our detector's bandwidth, and evaluate the concentrations $\{c_k\}$ from BL law (eq.~\ref{eq:BeerLambert}), using an appropriate optimization method. In this context, FTIR spectroscopy is typically limited by the overlap of absorption cross-sections of different gasses, low sensitivity, as well as fluctuations in the atmospheric conditions and the source's spectrum~\cite{FTIRlimitOpenPath,FTIRBook}. We leverage differential absorption spectroscopy (DAS) techniques in order to mitigate these limitations. Initially developed for detecting interfering pollutants in the ultraviolet-visible spectrum \cite{DOASBook}, DAS consists in decomposing the measured absorbance $\Tilde{\mathcal{A}}(\nu_i)$ (see eq.~\ref{eq:TransmEval}) into two terms:
\begin{equation}\label{eq:decompAbsorbance}
    \Tilde{\mathcal{A}}(\nu_i) = \overline{\mathcal{A}}(\nu_i) + \Delta\mathcal{A}(\nu_i),
\end{equation}
where $\overline{\mathcal{A}}(\nu_i)$ is a slow-varying term, and $\Delta\mathcal{A}(\nu_i)$ is the remaining part, called \textit{differential absorbance}, which features thin absorption lines from the probed gasses. Under reasonable assumptions on our experiments, it was shown that $\Delta\mathcal{A}(\nu_i)$ takes the following form \cite{DOASBook}:
\begin{equation}\label{eq:diffAbsorbance}
\Delta\mathcal{A}(\nu_i) = \overline{\mathcal{A}}(\nu_i)-     \Tilde{\mathcal{A}}(\nu_i) = L\sum_{k=1}^N c_k\bigl[\Delta\sigma_k*H\bigr](\nu_i),
\end{equation}
where $\Delta\sigma_k(\nu_i) =  \sigma_k(\nu_i) - \overline{\sigma}_k(\nu_i)$ are the \textit{differential cross-sections}, obtained in the same way as $\Delta\mathcal{A}(\nu_i)$ in order to erase the smooth terms $ \overline{\sigma}_k(\nu_i)$. In this way, deviations from the theoretical spectra, typically due to fluctuations in the source's spectrum and atmospheric conditions, are absorbed in $\overline{\mathcal{A}}(\nu_i)$ and ignored from the analysis. In addition, $\Delta\mathcal{A}(\nu_i)$ is now a linear function of ${c_k}$, contrary to the absorbance in eq.~\ref{eq:TransmEval} because of the detector's response $H$. The theoretical differential absorbance $\Delta\mathcal{A}(\nu_i)$ can therefore be used to fit the experimental data, via a fast linear least-square (LLS) method, which facilitates real-time analysis.\\

\begin{figure*}
    \centering
\includegraphics[width=155mm]{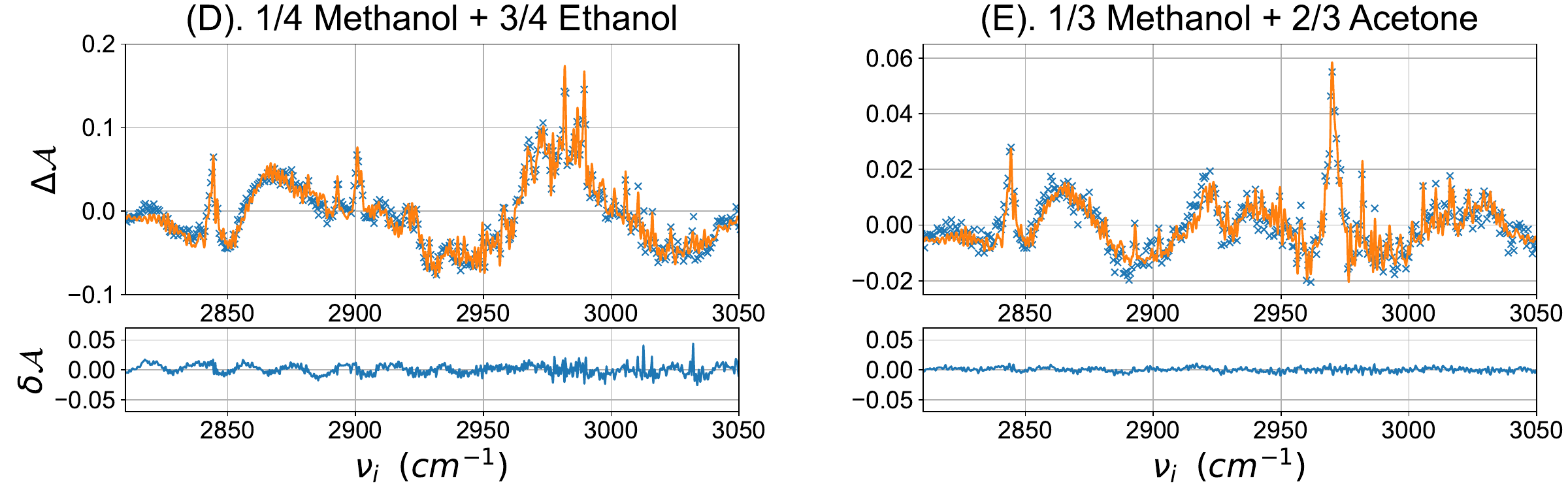}
    \caption{Differential absorbance $\Delta\mathcal{A}$ recorded by our QFTIR spectrometer, probing vapors emitted from liquid mixtures made of (D) 1/4 methanol and 3/4 ethanol, (E) 2/3 acetone and 1/3 methanol. The experimental data (in blue) is fitted by the theoretical absorbance (in~yellow, see expression \ref{eq:diffAbsorbance}). We also display the residual $\delta\mathcal{A}$.}
    \label{fig:2VOCs}
\end{figure*} 

The effectiveness of DAS relies on a few assumptions regarding the detector and gasses involved. First, for the expression \ref{eq:diffAbsorbance} to be valid, the detector's resolution $\Delta \nu$ should be very small compared to the spectrometer's bandwidth, the source's spectrum should be typically smooth, and the sample's absorption should be typically weak \cite{DOASBook}. All these assumptions are true in our experiments and also allow the approximation of eq.~\ref{eq:TransmEval}. Most importantly, as pointed out in \cite{IR-DOAS-thesis2004}, the gasses' cross-sections should feature specific narrow absorption lines within our detector's bandwidth. This is in particular the case for gaseous acetone (\ch{CH3COCH3}), methanol (\ch{CH3OH)}, and ethanol (\ch{CH3CH2OH}) \cite{Hitran2020}, 3 common VOCs present both in domestic and industrial contexts. Our spectrometer is therefore particularly suited for the detection and discrimination of these vapors by DAS, despite the overlap of their cross-sections.\\

In order to demonstrate the ability of our QFTIR spectrometer to detect these 3 VOCs, we first probed vapors emitted in ambient air from pure liquid acetone (A), methanol (B) and, ethanol (C), poured on the idler's arm of the interferometer. As the cross-sections of these 3 VOCs are typically lower than methane's, we increase the detector's sensitivity to the compounds by averaging the spectra over 500 scans instead of 100. We evaluate the absorbance $\Tilde{\mathcal{A}}(\nu_i)$ from eq.~\ref{eq:TransmEval}, by acquiring the reference spectra $I_0(\nu_i)$ in ambient air, with no liquid sample placed in the idler's arm. The slow varying term $\overline{\mathcal{A}}(\nu_i)$ from eq.~\ref{eq:decompAbsorbance} is subtracted by polynomial fit of $\Tilde{\mathcal{A}}(\nu_i)$, giving the differential absorbance $\Delta\mathcal{A}(\nu_i)$, displayed in Fig.~\ref{fig:CSVOC}. We proceed similarly with the cross-sections $\{\sigma_k\}_{k=1,2,3}$ found on the Hitran database \cite{Hitran2020}, which gives the differential cross-sections $\{\Delta\sigma_k\}_{k=1,2,3}$ from eq~\ref{eq:diffAbsorbance} (see supp. mat.~\ref{sect:DiffCS}). We then use this expression to fit the measured absorbance (see Fig.~\ref{fig:CSVOC}), with concentrations $\{c_k\}_{k=1,2,3}$ left as parameters. The resulting estimated average concentrations are given in the first 3 rows of Table~\ref{tab:data}. In each of these experiments, we estimate the noise as the standard deviation on the residual $\delta\mathcal{A}$, deviation of the measured $\Delta\mathcal{A}$ from the fit. In this way, we estimate SNRs of $\approx 51$ for experiment~A with acetone, $\approx 22$ for experiment~B with methanol, and $\approx 49$ for experiment~C with ethanol. We deduce the minimum average concentration detectable by our spectrometer, or detection limit, for each gas. We verify afterward that in each of these experiments, the average concentrations of absent gasses are lower than their detection limits, so the gasses are indeed not detected by our QFTIR spectrometer. Therefore, the probed gasses were properly identified.\\

\begin{figure}[htbp]
    \centering
\includegraphics[width=70mm]{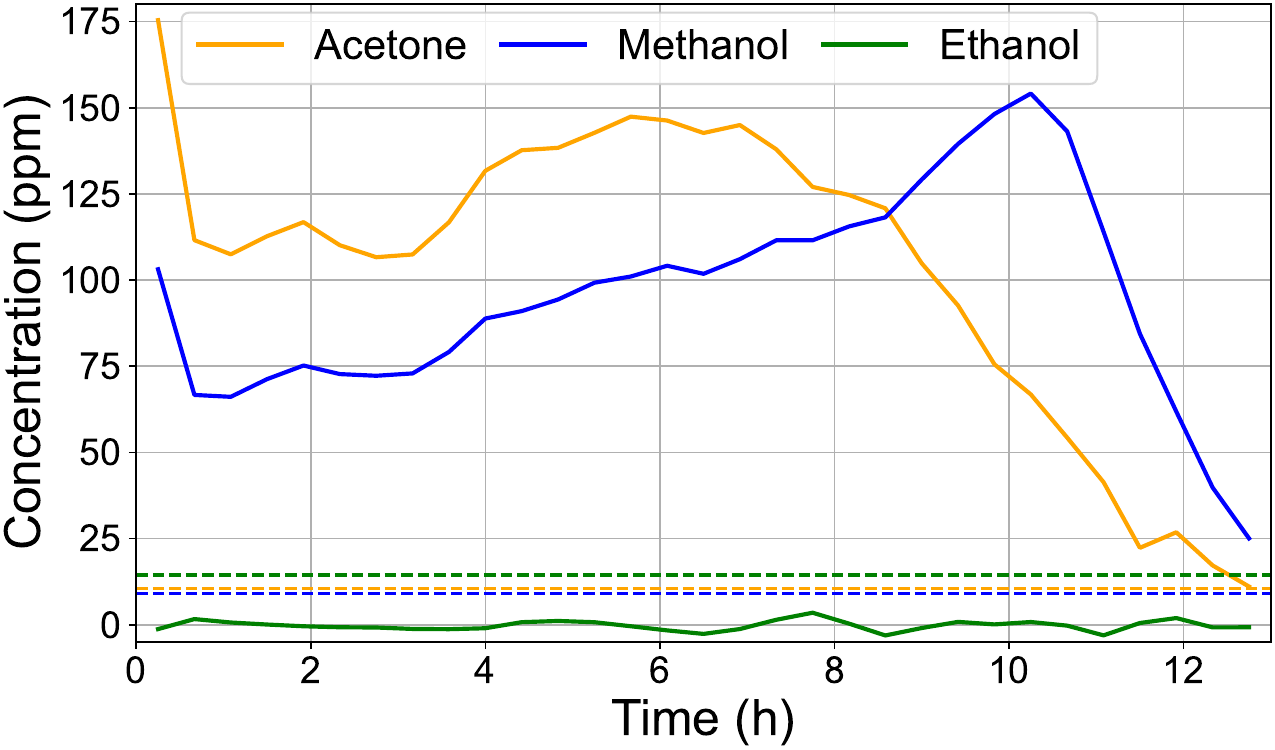}
    \caption{Evolution of the 3 VOCs' concentrations, estimated by analysing the acetone-methanol mixture during experiment~E, by QFTIR spectroscopy. Dashed lines show the detection limits of each gas.}
    \label{fig:Concentrations}
\end{figure}

We then performed 2 experiments to further demonstrate the ability of our detector to discriminate the VOCs in a mixture. We probed vapors emitted in ambient air from a liquid mixture with $1/4$ of ethanol and $3/4$ of ethanol (D), and from a liquid mixture with $2/3$ of acetone and $1/3$ of methanol (E). The same measurement and analysis were performed as for pure liquid mixtures, except experiment E was repeated multiple times overnight, in order to record the evolution of the absorbance and concentrations during the evaporation. The measured differential absorbance is displayed in Fig.~\ref{fig:2VOCs} together with the fit and residual, for experiment D as well as the first iteration of experiment E. The resulting estimated concentrations of the 3 VOCs are given in Table~\ref{tab:data}. We provide the absorbance evolution during the evaporation online \cite{gifXPE}, and the evolution of estimated concentrations during experiment E is displayed in Fig.~\ref{fig:Concentrations}. In each of these experiments, the average concentration of absent gas was below the detection limit, whereas concentrations of the present compounds were measured above that limit. Again, our detector was thus able to properly analyze the probed vapors, despite the interfering cross-sections of the compounds.

\def\arraystretch{1.5}
\begin{table}[htbp]
\centering
\begin{tabular}{| c || c | c | c |} 
 \hline
 Liquid Mixture  & \hspace{1mm}Acetone\hspace{1mm} & \hspace{1mm} Methanol\hspace{1mm} &\hspace{1mm}  Ethanol\hspace{1mm}  \\ \hline
 Pure Acetone & {\color{Green} 529} &  {\color{red} -6.89} &  {\color{red} -2.58} \\ 
 Pure Methanol &  {\color{red} -9.92} & {\color{Green} 197} &  {\color{red} -3.85} \\
 Pure Ethanol &  {\color{red} 4.66} &  {\color{red} -5.61} & {\color{Green} 707}\\
 \hline
 1/4 Methanol + 3/4 Ethanol &  {\color{red} $\SI{9e-3}{}$} & {\color{Green} 327} & {\color{Green} 381} \\
 2/3 Acetone + 1/3 Methanol  & {\color{Green} 175} & {\color{Green} 103} & {\color{red} -1.28} \\ 
 \hline\hline
 Detection limit & {\color{blue}10.4} & {\color{blue}8.95} & {\color{blue}14.4} \\
 \hline
\end{tabular}
\caption{Average concentrations of the 3 VOCs in the vapors emitted from the different liquid mixtures, estimated by QFTIR spectroscopy. Numbers are given in ppm, and displayed in green, resp. red, if they are above, resp. below, the detection limit given in the last row.}
\label{tab:data}
\end{table}
\medskip
\large
\noindent\textbf{Discussion}
\normalsize

We demonstrated the first open-path detection of multiple VOC vapors in ambient air, using QFTIR spectroscopy. By employing elongated interferometer arms, we achieved unprecedented sensitivity to methane for a QFTIR spectrometer, and characterized for the first time its sensitivity to 3 common VOCs: acetone, methanol, and ethanol.  Although rarely used in FTIR spectroscopy due to low sensitivity or narrow accessible bandwidth, DAS was leveraged for the first time with QFTIR spectroscopy. By leveraging this technique, we overcame limitations caused by fluctuations in ambient air and the source's spectrum, as well as inaccuracies resulting from the interfering cross-sections of the 3 compounds. The ability of accurately identifying pure or mixed VOC vapors was shown. This property, known as \textit{cross-selectivity}, is pivotal for multi-component chemical sensors \cite{CrossSelect}. Finally, we used our detector to analyze the vapors emitted by evaporation of an acetone-methanol mixture, demonstrating the first practical use-case of a QFTIR spectrometer as a VOC sensor. Our work therefore consolidates QFTIR spectroscopy as a promising candidate for the remote analysis of unknown gaseous mixtures in ambient air.\\

\large
\noindent\textbf{Acknowledgments}
\normalsize
\medskip

\noindent We acknowledge financial support from the NATO-SPS project HADES (id G5839), as well as Marco Barbieri and Ilaria Gianani from Roma Tre University, and Andrea Chiuri from ENEA, for fruitful and constructive conversations.\\

\large
\noindent\textbf{Data Availability}
\normalsize
\medskip

\noindent The data that support the findings of this study are available from the corresponding author upon reasonable request.\\

\large
\noindent\textbf{Author Contributions}
\normalsize
\medskip

G.G., L.L.V. and J.P.W designed the original experiment, and all authors contributed to the development of the setup. S.N. and A.K. optimized and characterized the interferometer. S.N. performed the experiments of VOC detection in ambient air, with the support of J.P.W.. S.N. analyzed and interpreted the data, with the support of A.K. and J.P.W.. S.N. wrote the manuscript, with the support of A.K., S.G. and J.P.W.. All authors reviewed the manuscript. J.P.W. supervised the project.

\onecolumngrid

\newpage
\appendix
\section*{Supplementary Material}

\section{Detector's Calibration With Methane}\label{sect:methane}

We calibrated our detector using a reference gas, made of a mixture of nitrogen $\ch{N2}$ (transparent in the MIR region) and $\SI{100}{ppm}$ of methane $\ch{CH4}$. We probed a $\SI{135}{\milli\meter}$ sealed gas-cell filled with $\SI{1}{bar}$ of this gas. By comparing the resulting spectrum with that obtained when probing the same cell filled with pure nitrogen, we get the absorbance shown in Fig.~\ref{fig:methane100}. We fit the data with eq.~\ref{eq:TransmEval}, using a nonlinear least-square method:
\begin{equation}\label{eq:TransmEval2}
   \Tilde{\mathcal{A}}(\nu_i)= -\ln\Bigl(\dfrac{[I_1 * H](\nu_i)}{[I_0 * H](\nu_i)}\Bigr) \simeq -\ln\bigl([T * H](\nu_i)\bigr),
\end{equation}
where $T$ is given by Beer-Lambert's law (see eq.~\ref{eq:BeerLambert}), and $H$ is the detector's response (see eq.~\ref{eq:detResp}). The methane concentration $c_{\ch{CH4}}$ and detector' resolution $\Delta\nu$ are left as free-parameters. We retrieve a concentration $c_{\ch{CH4}} = \SI{100.4}{ppm}$, and a resolution $\Delta\nu = \SI{1.11}{\per\cm} \simeq (\Delta x)^{-1}$, which are very close to the expected values. We measure a SNR of $\simeq 24$ for the central peak around $\SI{3018}{\per\cm}$, giving a detection limit of $\lesssim\SI{5}{ppm}$ for methane in the $\SI{1.35}{\meter}$ gass-cell, or of $\lesssim\SI{4}{ppm}$ for methane spread over the $\lesssim\SI{1.7}{\meter}$ interferometer-arm.\\

\medskip

\begin{minipage}{0.50\textwidth}
    \centering
\includegraphics[width=89mm]{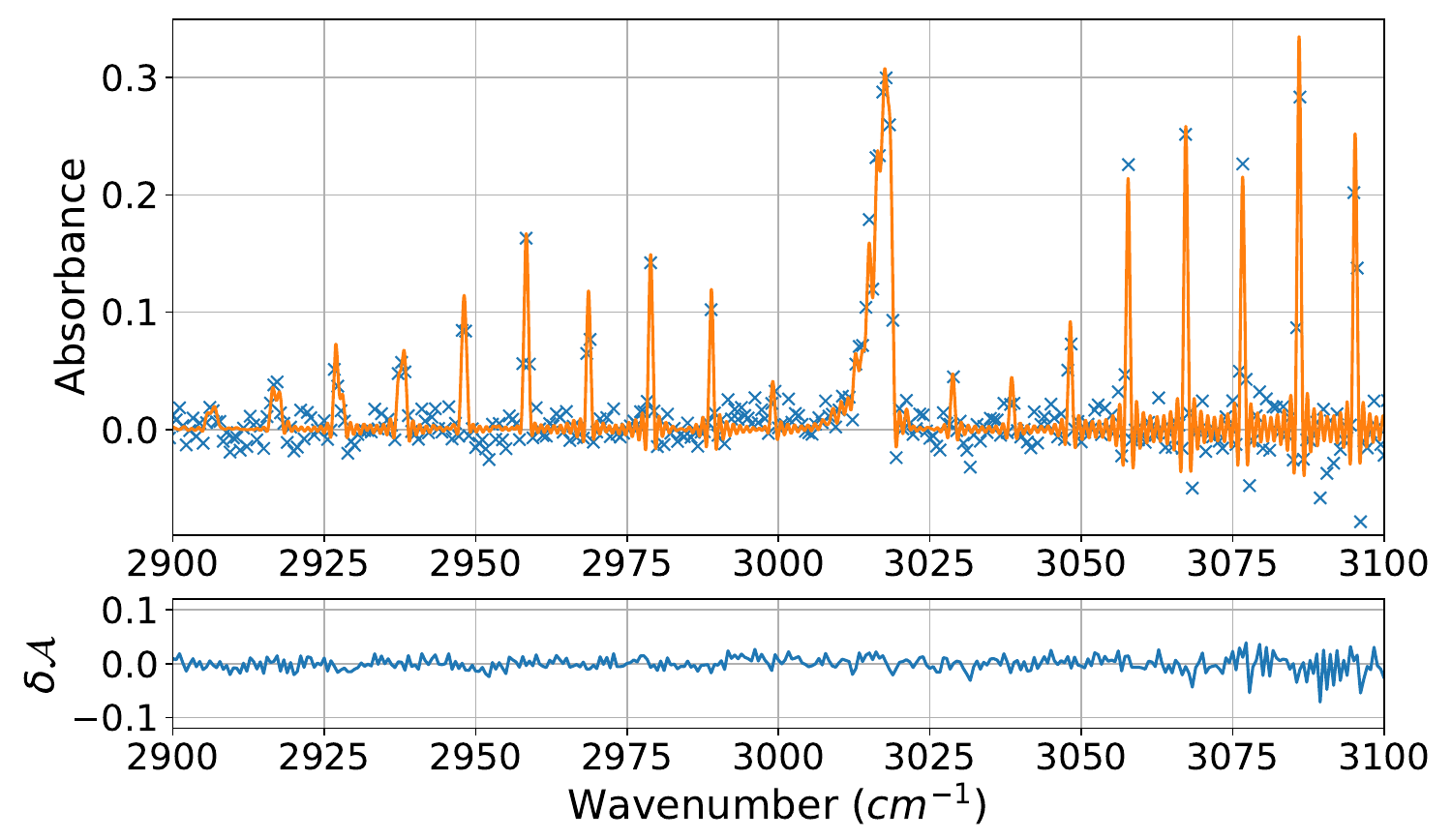}
    \captionof{figure}{Absorbance recorded by probing the $\SI{1.35}{\meter}$ gas-cell, filled with $\SI{100}{ppm}$ of methane in a $\SI{1}{\bar}$ mixture with nitrogen. The orange line displays the fit with eq.~\ref{eq:TransmEval2}, obtained by nonlinear least-square method.}
    \label{fig:methane100}
\end{minipage}
\hspace{0.5cm}
\begin{minipage}{0.44\textwidth}
    \centering
\includegraphics[width=75mm]{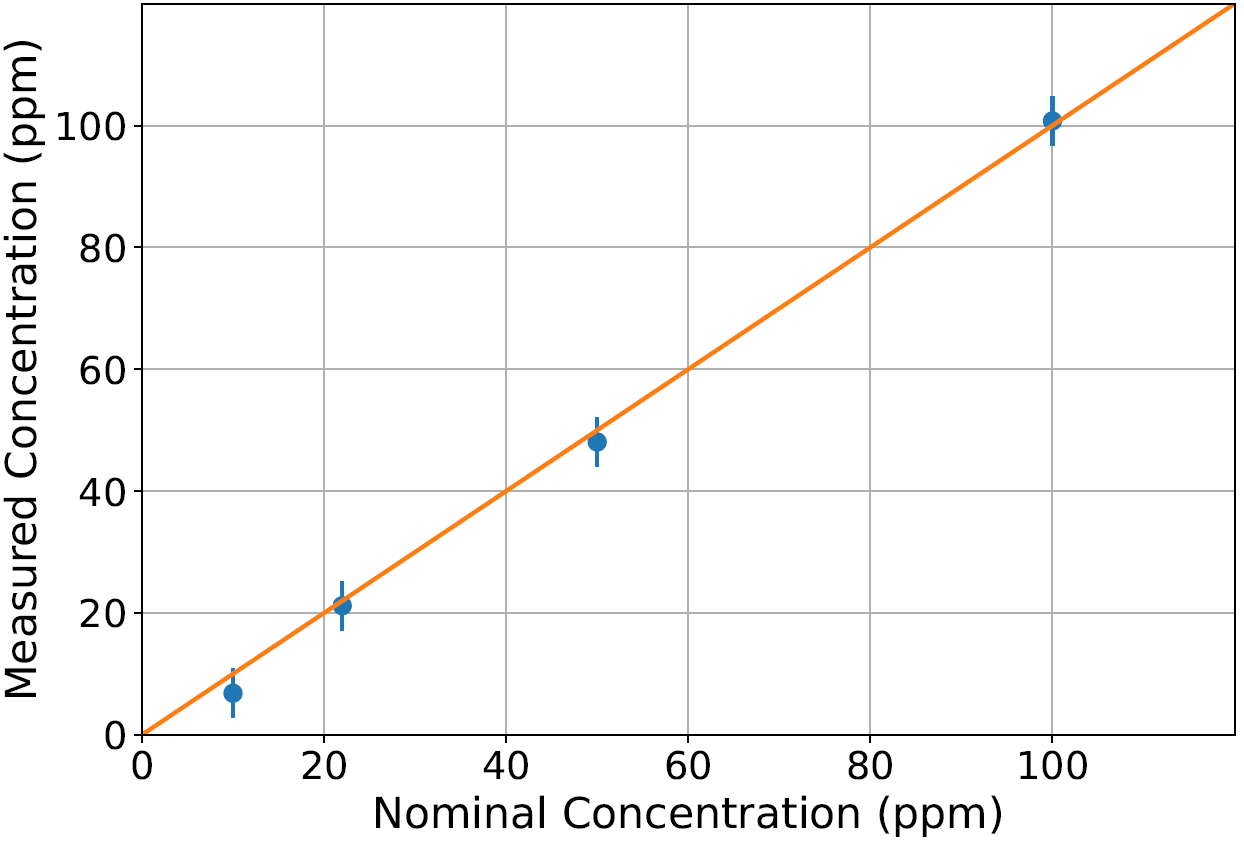}
    \captionof{figure}{Methane concentrations in the $\SI{135}{\milli\meter}$ sealed gas cell, estimated via our detector, versus the nominal value. The error bars are given by the $\SI{5}{ppm}$ detection limit, and the orange line is the identity function.}
    \label{fig:methaneConcentrations}
\end{minipage}

\medskip
\medskip

Finally, we reproduced the same experiment with different methane concentrations, by mixing the reference gas with nitrogen. In this way, we probed mixtures with $\SI{50}{ppm}$, $\SI{22}{ppm}$ and $\SI{10}{ppm}$ of methane. By the same fitting procedure, we retrieve the nominal concentrations within the $\pm\SI{5}{ppm}$ interval, given by the detector's detection limit. These concentrations are displayed in Fig.~\ref{fig:methaneConcentrations}. This further shows the ability of our spectrometer to accurately measure gas concentrations.

\newpage

\section{Differential Cross-Sections}\label{sect:DiffCS}

\begin{figure}[htbp]
    \centering
\includegraphics[width=179mm]{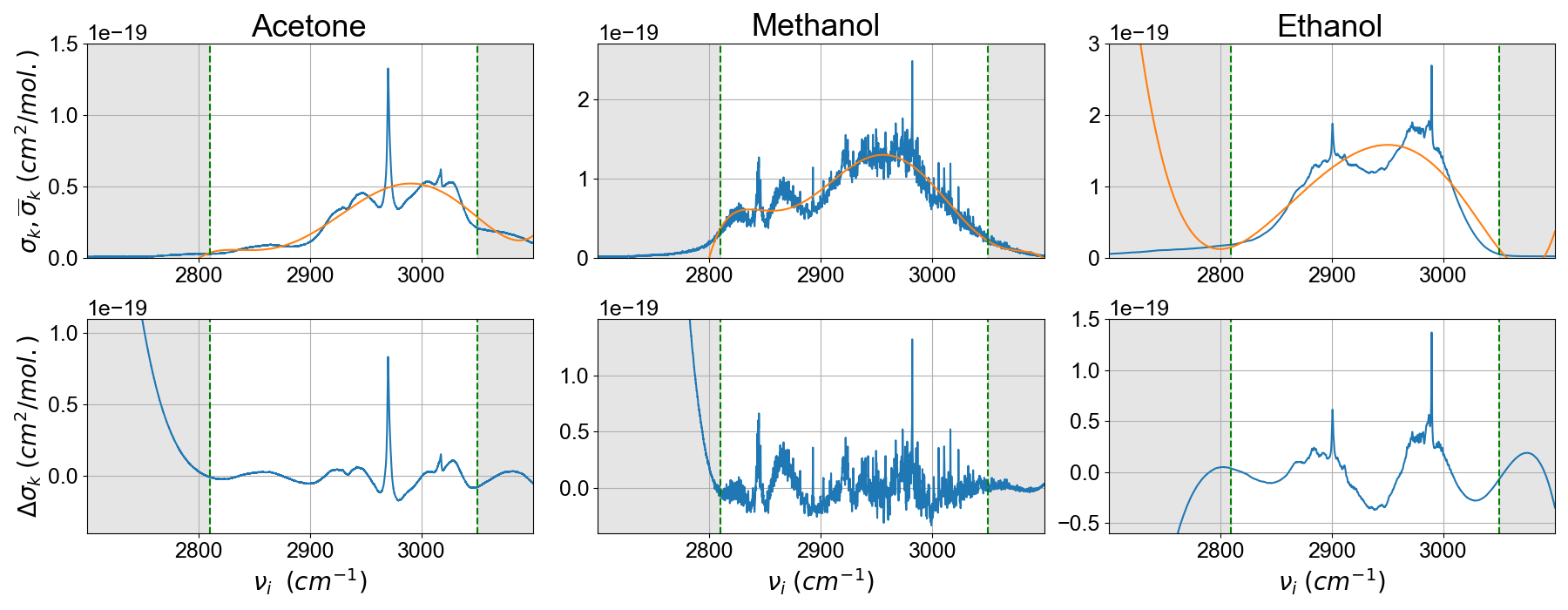}
    \caption{First line: Absolute cross-sections $\sigma_k(\nu_i)$ of the 3 VOCs (in blue), and the polynomial fit giving the slow-varying terms $\overline{\sigma}_k(\nu_i)$ (in orange). Second line: Differential cross-sections $\Delta\sigma_k(\nu_i) = \sigma_k(\nu_i) - \overline{\sigma}_k(\nu_i)$ of the corresponding VOCs. The dashed green lines show the limits of the relevant spectral region, outside of which the differential cross-sections diverge.}
    \label{fig:diffCSVOCs}
\end{figure}

The key step for implementing DAS is to compute the differential cross-sections $\{\Delta\sigma_k\}$ from the absolute cross-sections $\{\sigma_k\}$, found on the HITRAN database \cite{Hitran2020}. As suggested in \cite{DOASBook}, the slow-varying terms $\{\overline{\sigma}_k\}$ can be removed by polynomial fit of $\{\sigma_k\}$. In our case, we use polynomials of degree 9, accounting for numerous slow inflections of the cross-sections over our detector's bandwidth, as well as fluctuations in the experimental data. We restrict the fit to the bandwidth between $\SI{2810}{\per\cm}$ and $\SI{3050}{\per\cm}$, which include all high-absorption peaks, and outside of which the cross-sections vanish. Note that the same procedure is carried out in order to obtain the differential absorbance $\Delta\mathcal{A}(\nu_i)$ from the measured absorbance $\Tilde{\mathcal{A}}(\nu_i)$. In Fig.~\ref{fig:diffCSVOCs}, we display the cross-sections $\{\sigma_k\}$ taken from the Hitran database for the 3 VOCs of interest, the slow-varying terms $\{\overline{\sigma}_k\}$ obtained by polynomial fit, and the resulting differential cross-sections $\{\Delta\sigma_k\}$ used in eq.~\ref{eq:diffAbsorbance} to fit the experimental differential absorbance.\\

\end{document}